\documentclass[fleqn,usenatbib]{mnras}

\usepackage[T1]{fontenc}
\usepackage{ae,aecompl}

\usepackage{graphicx, float}	
\usepackage{amsmath}	
\usepackage{amssymb}	

\title[Variation of coronal temperature with flux]{{\it NuSTAR} observation of Ark 564 reveals the variation of coronal temperature with flux}

\author[Barua et al.]{
Samuzal Barua$^{1}$\thanks{E-mail: samuzal.barua@gmail.com},
V. Jithesh$^{2}$\thanks{E-mail: vjithesh@iucaa.in},
Ranjeev Misra$^{2}$\thanks{E-mail: rmisra@iucaa.in},
Gulab C Dewangan$^{2}$,
\newauthor 
Rathin Sarma$^{3}$ and Amit Pathak$^{4}$ \\
$^{1}$Department of Physics, Gauhati University, Jalukbari, Guwahati-781014, Assam, India\\
$^{2}$Inter-University Centre for Astronomy and Astrophysics (IUCAA), PB No.4, Ganeshkhind, Pune-411007, India\\
$^{3}$Department of Physics, Rabindranath Tagore University, Hojai-782435, Assam, India\\
$^{4}$Department of Physics, Banaras Hindu University, Varanasi-221005, India\\ 
}

\date{Accepted XXX. Received YYY; in original form ZZZ}

\pubyear{2020}

\begin{document}
\label{firstpage}
\pagerange{\pageref{firstpage}--\pageref{lastpage}}
\maketitle

\begin{abstract}
The hard X-ray spectral index of some AGN has been observed to steepen with the source flux. This has been interpreted in a Comptonization scenario, where an increase in the soft flux decreases the temperature of the corona, leading to steepening of the photon index. However, the variation of the coronal temperature with flux has been difficult to measure due to the presence of complex reflection component in the hard X-rays and the lack of high-quality data at that energy band. Recently, a 200 ks {\it NuSTAR} observation of Ark 564 in 3--50 keV band, revealed the presence of one of the coolest coronae with temperature $kT_e \sim 15$ keV in the time-averaged spectrum. Here, we re-analyse the data and examined the spectra in four flux levels. Our analysis shows that the coronal temperature decreased from $\sim 17$ to $\sim 14$ keV as the flux increased. The high energy photon index $\Gamma \sim 2.3$ varied by less than $0.1$, implying that the optical depth of the corona increased by about 10\% as the flux increased. This first reporting of coronal temperature variation with flux shows that further long observation by {\it NuSTAR} of this and other sources would shed light on the geometry and dynamics of the inner regions of the accretion flow.

\end{abstract}

\begin{keywords}
black hole physics -- galaxies: active -- galaxies: Seyfert -- X-rays: individual: Ark 564
\end{keywords}



\section{Introduction}

An important characteristic Narrow-line Seyfert 1 (NLS1) galaxies 
is  that they typically show extreme variability
in the X-ray band \citep{1991Natur.350..589R, 1997MNRAS.289..393B, Brandt1998, 1999MNRAS.303L..53B}. It is believed that they
harbour smaller black holes and accrete at near Eddington limit \citep{1995MNRAS.277L...5P, Collin2004} compared to standard Active Galactic Nuclei (AGN).
NLS1s generally show strong soft X-ray excess below 2 keV, and their spectral slopes are somewhat steeper \citep{Boller1996, Brandt1997, Turner1998, Leighly1999b, Vaughan1999b}.

It is nearly a standard paradigm, that the X-ray power law emission of AGN, is due to inverse-Compton
scattering of soft photons from a standard accretion disc, by the electrons of an
hot corona \citep{1979Natur.279..506S}. The Comptonized 
spectra from the hot thermal corona impinge on the standard disk resulting in a 
reflected component detected by a  fluorescent iron line peaking at 
$\sim 6.4$ keV and a Compton reflection hump at 20--100 keV \citep{George1991}. 
One can fit broadband X-ray data with a thermal Comptonization model and
corresponding reflection component to obtain an estimate of the coronal
temperature. However, till recently, it was difficult to do so because of
the low sensitivity of detectors in this energy range and the complexities
arising from modelling the reflection component.


{\it The Nuclear Spectroscopic Telescope Array (NuSTAR)} is the first
hard X-ray focusing telescope in orbit \citep{Harrison2013}, which
operates in the 3--79 keV X-ray band, extending the sensitivity of
focusing optics far beyond the $\sim$10 keV high-energy cutoff achieved by
all previous X-ray satellites.  \citet{Gar15} have shown
that with these specifications, it is possible to measure the temperature
(or the high energy cutoff) of AGN coronae from spectral analysis of {\it NuSTAR} data. Indeed a long observation of Ark 564 by {\it NuSTAR} revealed the
presence of a low temperature corona \citep[$\rm kT_{e} \sim 15$ keV;][hereafter EK17]{Kara2017}. Recently, {\it NuSTAR} observations
provided estimates of low temperatures ($\rm kT_{e} \sim 20$ keV) of the corona for
two more AGN \citep{Buisson1093}.

As mentioned earlier, AGN in general, and NLS1s, in particular, are highly
variable in both flux and spectral shape and the correlation between them,
provide insight into their nature. One of the interesting variable property,
they share with X-ray binaries is that typically the high energy photon index
correlates positively with the low energy flux \citep[for e.g.][]{Haardt1997,Zdziarski2003,Sobolewska2009,sarma2015}.
In other words, the high energy spectrum steepens with increasing soft flux. 
It has been argued that this may be a natural consequence of the
Comptonization \citep{Zdziarski2003}. If the flux of the input 
seed photons (i.e. the soft photons) increases, the radiative 
cooling of the corona increases. Since the radiative cooling 
efficiency is very high there must be heating of the corona to 
maintain a high temperature \citep{Fab15}. If this 
heating rate does not vary (or if it does not increase in the 
same proportion as the input flux) this would lead to a decrease 
in the temperature of the corona. This in turn would cause high 
energy spectra to steepen as observed. 
Although this interpretation is attractive and seems natural, it should be noted that up till now
there has been no direct evidence for the coronal temperature to decrease with flux. There are indirect ways to infer that the temperature decreases with
flux. For example for Mrk 335, \citet{2015MNRAS.449..129W} measured the
expansion of the corona with flux, which when combined with the change
in photon index, implied that the coronal temperature must have
decreased.


The long ($\sim 200$ ks) {\it NuSTAR} observation of Ark 564, where the coronal temperature was found
to be $\sim 15$ keV (EK17) in the time averaged spectrum, provides a unique opportunity to test the hypothesis of 
the coronal temperature varying with flux. Ark 564 is a X-ray bright NLS1 galaxy at a red-shift z =
0.02469 \citep{Huchra1999} with a 2--10 keV flux of $\sim 2 \times 10^{-11}$ erg cm$^{-2}$ s$^{-1}$ \citep{Vaughan1999} and the
corresponding luminosity is $\sim 2.4 \times 10^{43}$ erg s$^{-1}$ \citep{Turner2001}.
The source shows rapid variability, a very steep spectrum in the 0.3--10 keV band,
strong soft excess and iron K emission line \citep{1999ApJ...526...52T, Leighly1999, Turner2001}. Analysis of a number of {\it XMM-Newton} observations suggest that this high-Eddington NLS1 exhibits a prominent correlation between spectral slope and
flux \citep{sarma2015}. EK17 report that during the {\it NuSTAR} observation, the source exhibited a 
dramatic flare, in which the hard emission is delayed with respect to the soft one. Indeed the flux of the source varied
by about a factor of two, and hence it should be possible to verify whether the coronal temperature also varied.

Here, we re-analyse the 200 ks {\it NuSTAR} observation to investigate whether any correlation or trend exists between the coronal electron temperature and flux of the X-ray luminous source Ark 564. The paper is organized as follows: In \S 2, we describe the observations and analysis. In \S 3, we present the results of the spectral fitting analyses of time-averaged as well as the flux resolved spectra and discuss the interpretations of these results in \S 4.

\section{Observation and data analysis}

\begin{table*}
	\centering
	\caption{Spectral parameters from the simultaneous fits of {\it NuSTAR} FPMA and FPMB time-averaged spectra in 3--50 keV band using model {\tt xillverCp}. From left to right the model parameters are photon index ($\Gamma$), iron abundance (Afe), temperature of corona electrons ($\rm kT_{e}$), log of the ionization parameter or accretion disk ionization ($\xi$), inclination of the accretion disk ($\theta$), reflection fraction (R), normalization, flux ( $\rm F_{3-50\,keV}$) and reduced $\chi^2$ with degrees of freedom respectively.}
	\label{tab:Table1_table}
	\begin{tabular}{lcccccccc} 
		\hline
		 $\Gamma$ & Afe &  $\rm kT_{e}$  & log$\xi$ & $\theta$  & R & Norm & $\rm F_{3-50\,keV}$  & $\chi^2_{r}$\\
	              & solar& keV  & log(erg cm s$^{-1}$)    &   degree  &   & 10$^{-4}$& $\rm 10^{-11} erg\, cm^{-2}\, s^{-1}$   & /d.o.f     \\
		\hline

		 2.30$^{+0.040}_{-0.001}$ & 2.24$^{+0.10}_{-0.45}$ &14.97$^{+1.89}_{-0.79}$ & 4.65$^{+0.01}_{-0.23}$ & 30.97$^{+7.06}_{-3.76}$ & 1.23$^{+1.30}_{-0.65}$ & 0.45$^{+0.36}_{-0.19}$&1.93$^{+0.01}_{-0.01}$   & 1.02/619                      \\

		\hline
	\end{tabular}
\end{table*} 

Ark 564 was observed with {\it NuSTAR} on 2015 May 22 for an exposure
time of $\sim$ 200 ks. The {\it NuSTAR} data were reduced and cleaned
for high background flare with the {\sc nupipeline} script using {\it NuSTAR} data analysis software ({\sc nustardas}) v1.8.0 and CALDB version
20171002 to obtain calibrated event files in Level 1 data products. We
used a circular source region of 50 arcsec radius and a slightly
larger background region of 60 arcsec for both the detectors, Focal Plane Modules A and B (FPMA and FPMB).
With {\sc nuproducts} task, we extracted FPMA/FPMB spectra and light curves together
with the redistribution matrix files (RMF) and ancillary response
files (ARF). We grouped FPMA and FPMB time-averaged spectra using
{\sc grppha} tool with 50 counts in each spectral bin. Additionally as advised by \citet{Gar15}, we
binned the source spectra, background spectra and RMF in order to acquire the proper sampling of the data, 
which originally had equal bin width of 0.04 keV throughout the entire
X-ray band. In particular, we used a binning of 0.16 keV for data below 20 keV and a binning of 0.32 keV in the 20--50
keV band. 

\section{Spectral results}

\subsection{Time-averaged spectrum}

\begin{figure*}
	\centering
	\includegraphics[width=0.82\textwidth]{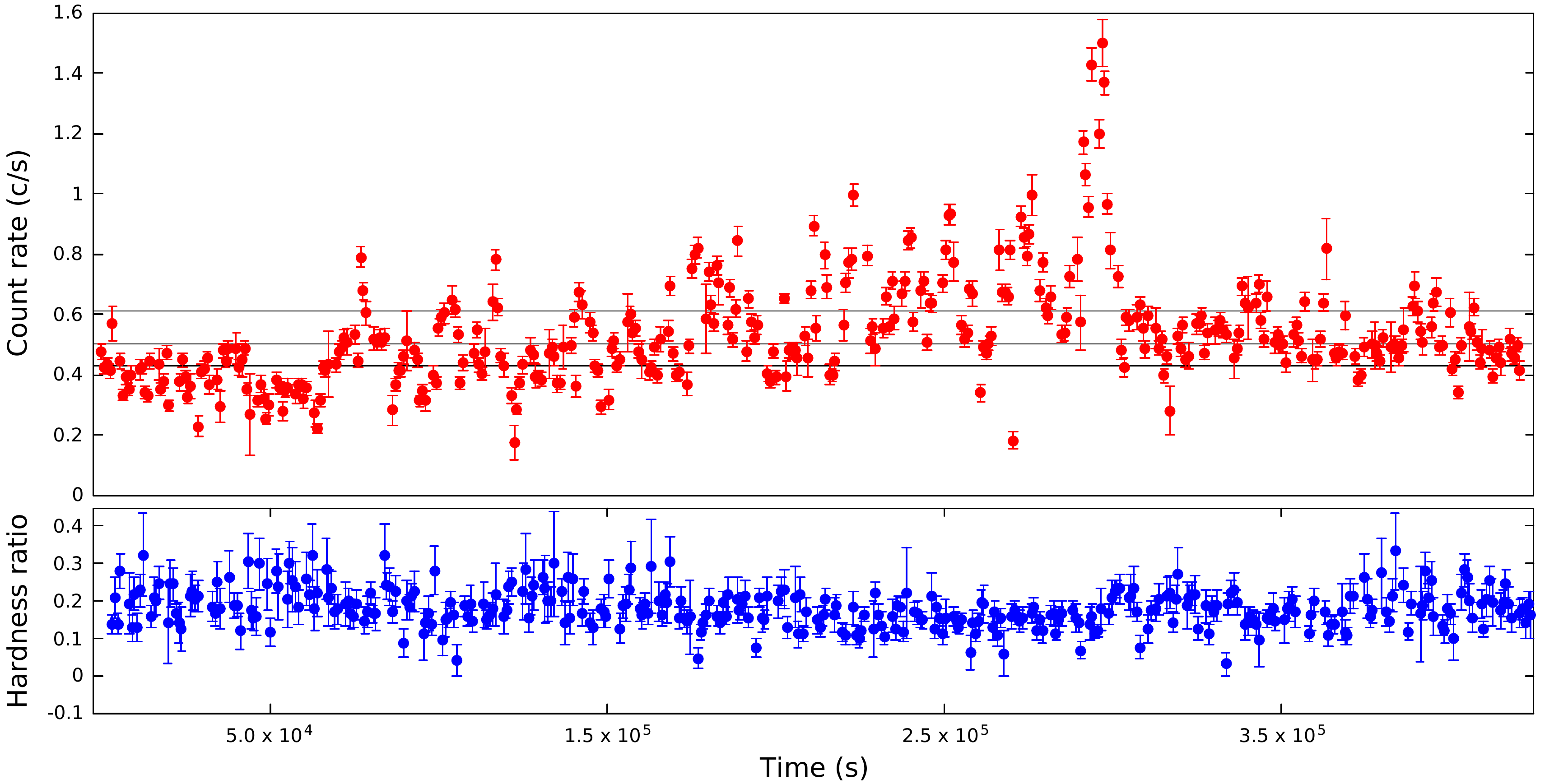}
	\caption{The light curve of Ark 564 from the {\it NuSTAR} observation with a total exposure time of 200 ks. The light curve is produced with a time bin size of 800 s. The black horizontal lines divide the light curve into four flux levels. The hardness ratio, depicted in the bottom panel, derived from the 3--10 keV and 10--50 keV energy bands.}
	\label{fig:lightcurve_figure}
\end{figure*}

We first reproduce the spectral results obtained by EK17. For that, we fitted the time-averaged spectrum from both FPMA
and FPMB detectors simultaneously in the 3--50 keV energy band using
the ionized reflection model {\tt xillverCp}\footnote{\url{http://www.sternwarte.uni-erlangen.de/~dauser/research/relxill/}},
where the continuum is modelled as being due to Comptonization \citep[i.e. the {\sc xspec} \citep{Arn96} model {\tt nthComp};][]{Zdziarski1996,zycki1999} and the corresponding ionized reflection.
We allowed for the possibility that relative normalization between the two instruments may vary,
by including a constant to the model for the FPMB spectrum. Errors on best-fit parameters were obtained from
the Markov chain Monte Carlo (MCMC) analysis by using the {\sc xspec\_emcee}
code\footnote{\url{https://github.com/jeremysanders/xspec_emcee}}. We
use the Goodman-Weare algorithm by specifying 60 walkers with 12,000
iterations to calculate the best-fit parameters and their uncertainty
(see more details in EK17). The best-fit parameters with errors from our analysis of the 
time-averaged spectrum are listed in Table 1. The best-fit parameter 
values correspond to the peaks of the marginalized probability 
distribution. The quoted parameter errors are at a 90\% confidence level.
The reported spectral parameters in EK17
such as electron temperature, $\rm kT_{e} = 15_{-1}^{+2}$ keV, the
photon index $\Gamma = 2.32_{-0.01}^{+0.02}$, reflection fraction
$R=1.18^{+2.10}_{-1.18}$ and disc ionization parameter, log$\xi = 4.43
\pm 0.06$ are consistent with our analysis, but other parameters are
marginally different from our results. The minor deviation in the
parameters may be due to the use of older versions of {\tt xillverCp}
model, {\it NuSTAR} software package (version 1.6.0) and CALDB
(version 20160502) in EK17 as compared to this analysis.  

Having obtained the spectral parameters similar to those obtained by EK17, we proceed to do flux resolved
spectroscopy to investigate if the coronal temperature shows any variation with flux.

\begin{table*}
	\centering
	\caption{Spectral Parameters from the flux resolved spectroscopy of Ark 564 for the four flux levels. In all spectral fits, the inclination ($\theta$) and iron abundance (Afe) parameters are fixed at the values obtained from the time-averaged spectral fit.}
	\label{tab:Table2_table}
	\begin{tabular}{lcccccccccc} 
		\hline
		Flux & $\Gamma$& &   $\rm kT_{e}$  && log$\xi$ &  R && Norm & $\rm F_{3-50\,keV}$ & $\chi^2_{r}$\\
		State  &   &&  keV    &&log(erg cm s$^{-1}$) &   && 10$^{-4}$   & $\rm 10^{-11} erg\, cm^{-2}\, s^{-1}$ & /d.o.f   \\
		\hline

		\vspace{0.2cm}
		1  & 2.31$^{+0.04}_{-0.05}$ &&17.41$^{+0.92}_{-0.67}$ && 4.44$^{+0.14}_{-0.16}$ & 1.22$^{+1.27}_{-1.17}$ && 0.37$^{+0.02}_{-0.01}$   & 1.62$^{+0.01}_{-0.01}$ & 1.11/416 \\
		\vspace{0.2cm}
		2  & 2.32$^{+0.04}_{-0.04}$ &&17.09$^{+0.91}_{-0.77}$ && 4.32$^{+0.21}_{-0.12}$ & 1.10$^{+1.14}_{-1.05}$ && 0.47$^{+0.02}_{-0.02}$   & 1.78$^{+0.01}_{-0.01}$ & 1.06/316 \\
		\vspace{0.2cm}
		3  & 2.32$^{+0.05}_{-0.05}$ &&16.04$^{+0.77}_{-0.80}$ && 4.60$^{+0.21}_{-0.24}$ & 1.12$^{+1.18}_{-1.07}$ && 0.49$^{+0.03}_{-0.02}$   & 1.95$^{+0.01}_{-0.01}$ & 1.18/331 \\
		4  & 2.35$^{+0.08}_{-0.08}$ &&14.01$^{+0.56}_{-1.06}$ && 4.69$^{+0.30}_{-0.27}$ & 1.12$^{+1.19}_{-1.06}$ && 0.79$^{+0.04}_{-0.05}$   & 2.82$^{+0.02}_{-0.02}$ & 1.01/365 \\
		\hline
	\end{tabular}
\end{table*}


\subsection{Flux-resolved spectroscopy} 

To extract the spectra for different flux states, the
FPMA and FPMB light curves were generated with a time bin of 800 s in the complete energy
range of 3--79 keV. The two light curves were then added and the resultant variation of count rate with time
is shown in Figure \ref{fig:lightcurve_figure}. A flare has
been detected during the observation of Ark 564, which is shown in the
light curve. The 3--79 keV mean count rates of the source in FPMA and FPMB are 0.28 c/s and 0.27 c/s, respectively. In order to investigate
the flux resolved spectra of Ark 564, the light curve is split into
four flux levels. The light curve is divided in such a way that each
level contains an approximately equal number of bins,
i.e., roughly 25 percent of the total number of bins. The four levels are
designated as very low ($< 0.42$ c/s), low (between 0.42 and 0.5 c/s), high (between 0.5 and 0.61 c/s) and very high ($> 0.61$ c/s) are shown in Figure \ref{fig:lightcurve_figure}. The
source spectra for both FPMA and FPMB along with the associated files
are extracted in a similar way as explained for the time-averaged
spectra.

\begin{figure}
	\centering
	\includegraphics[width=0.45\textwidth]{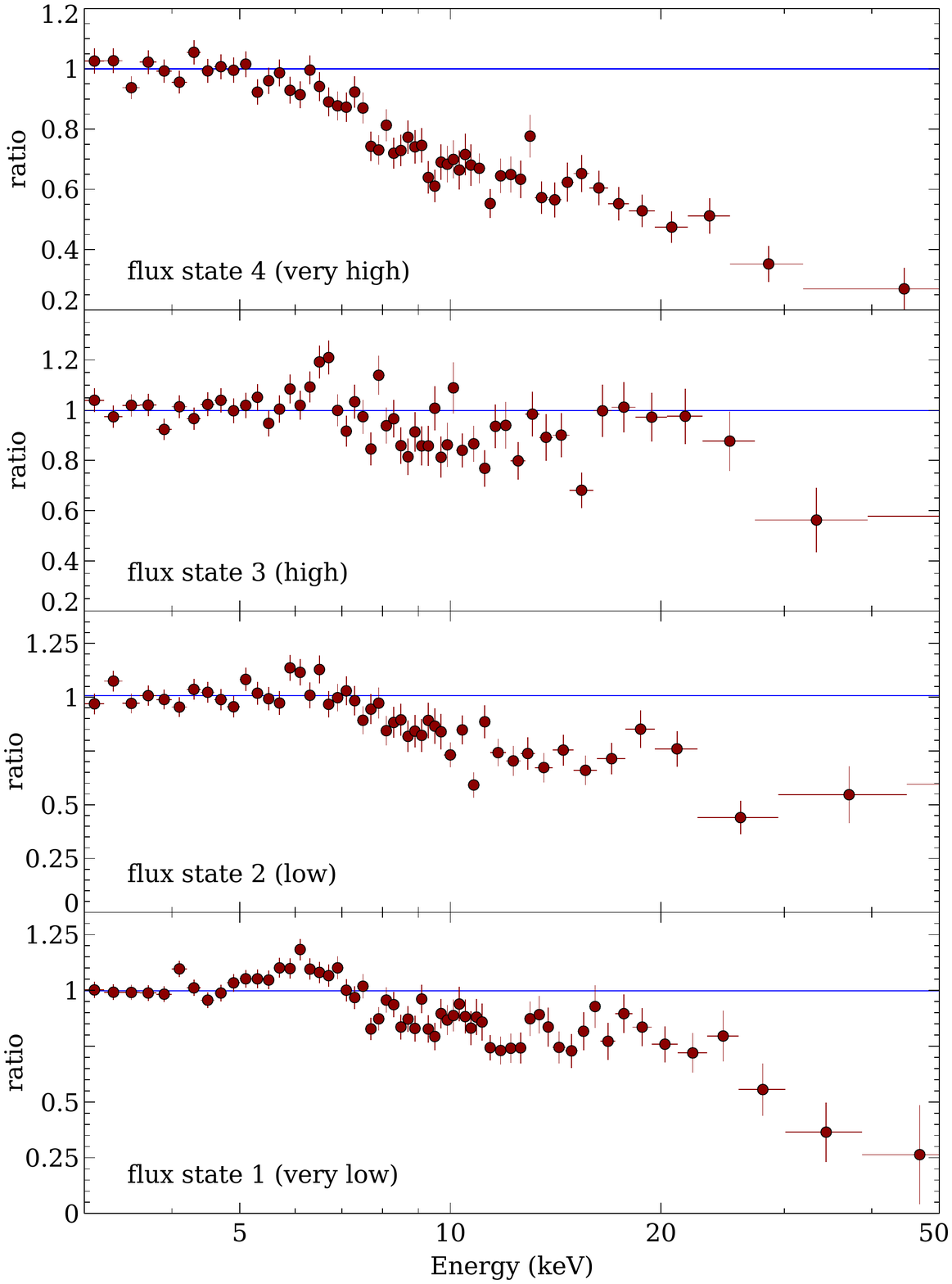}

	\caption{Data-model ratio from the power law fits of four {\it NuSTAR} FPMA flux resolved spectra. All of the FPMA spectra are grouped, and a standard binning is applied using {\sc xspec}.} 
	\label{Fig:ratio}
\end{figure}

To quantify the spectral shape difference between the four flux states in
a model-independent way, we fit the spectra in the energy range $3$-$5$ keV,
with a power-law and then show the ratio of the data to the model for the
full 3-50 keV range in Figure \ref{Fig:ratio}. While high energy curvature is seen in all
states, it is largest for the highest flux state, indicating perhaps that the
coronal temperature is lowest at that state.

\begin{figure}
	\includegraphics[width=0.57\textwidth,angle=0]{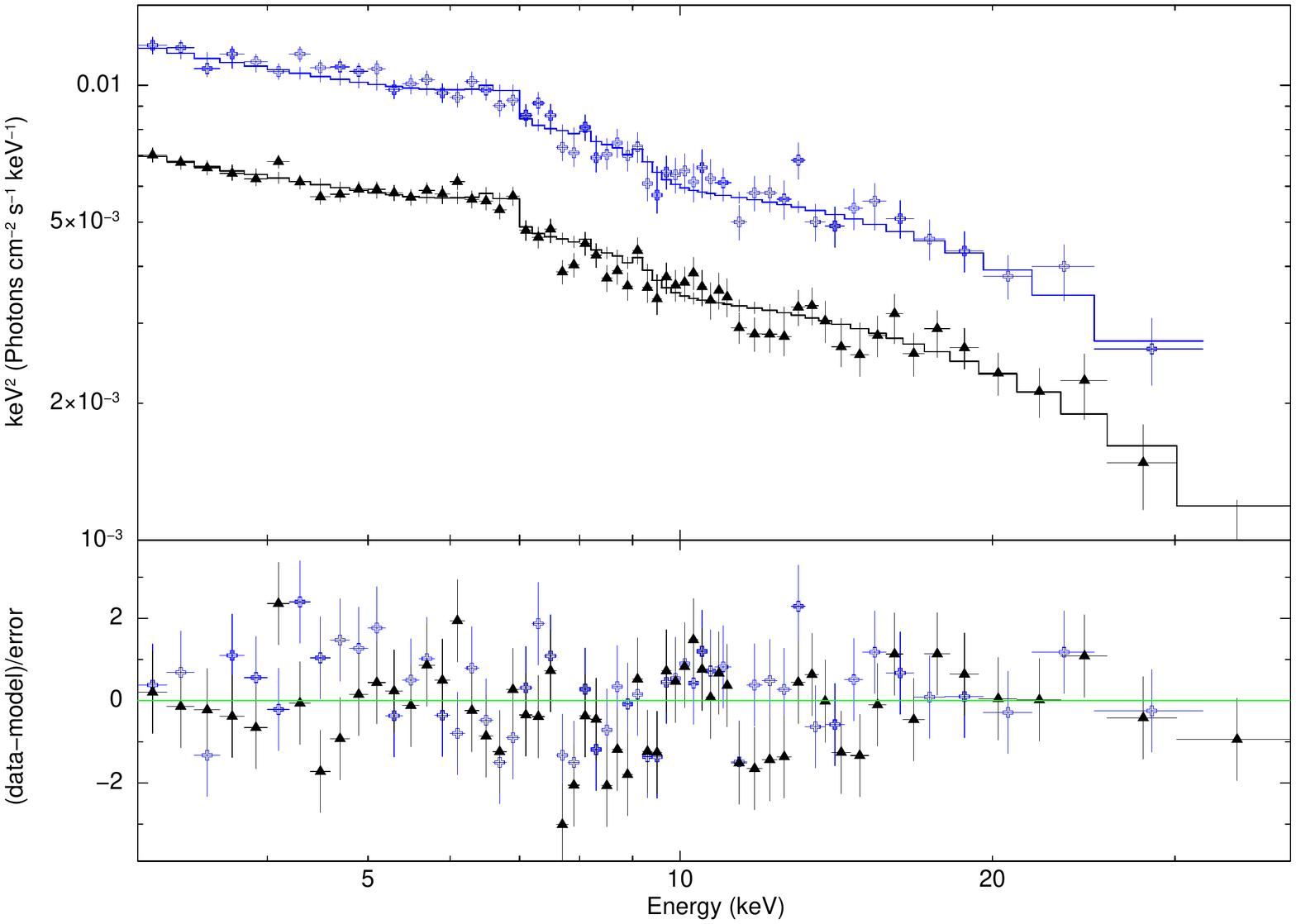}
	\caption{The NuSTAR FPMA spectra from the lowest (black triangle) and highest (blue plus) flux states fitted with xillverCp. All 3--50 keV spectra are grouped and binned as mentioned in \S 2.}
	\label{fig:4spectra}
\end{figure}

\begin{figure}
	\includegraphics[width=0.45\textwidth]{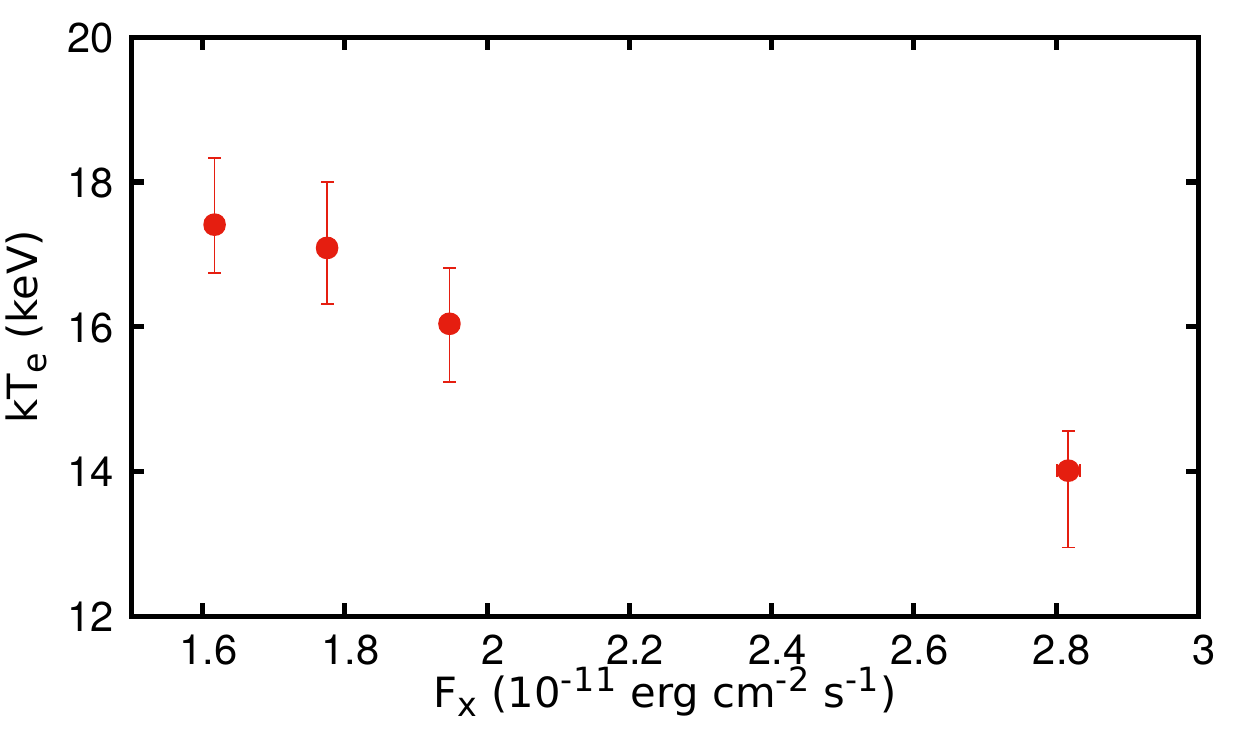}
	\includegraphics[width=0.45\textwidth]{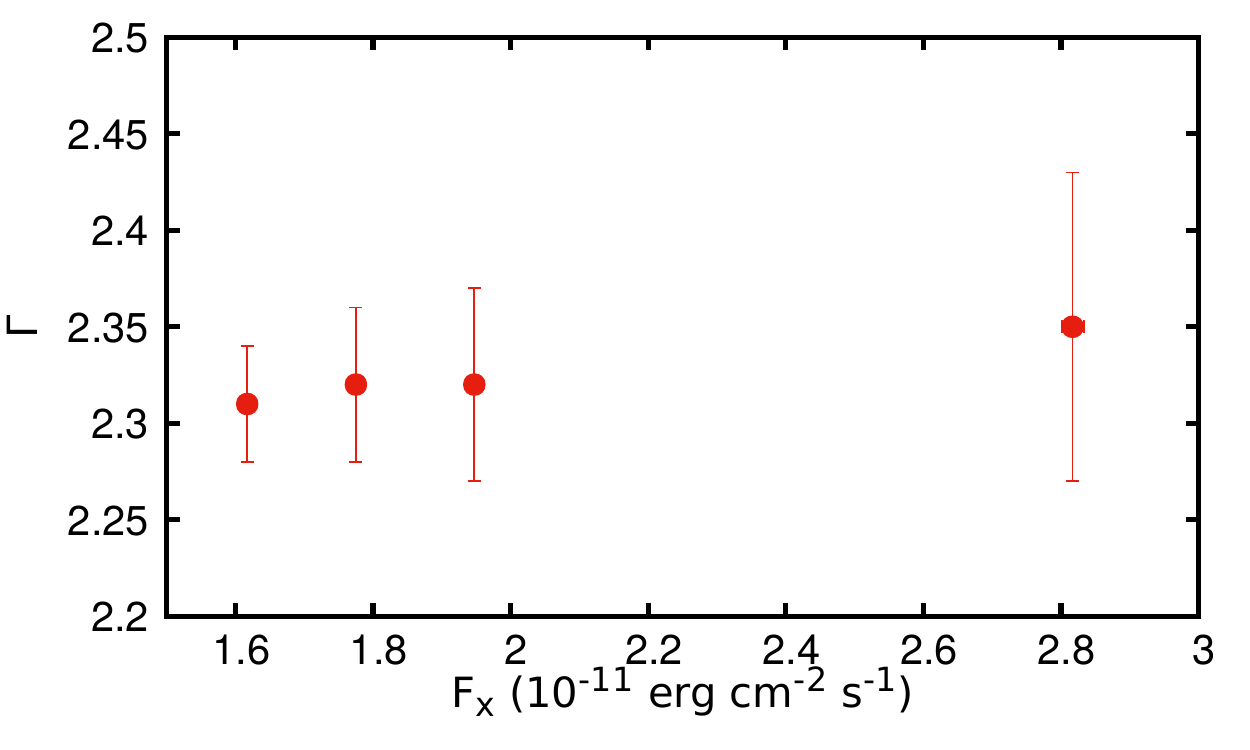}
	\includegraphics[width=0.45\textwidth]{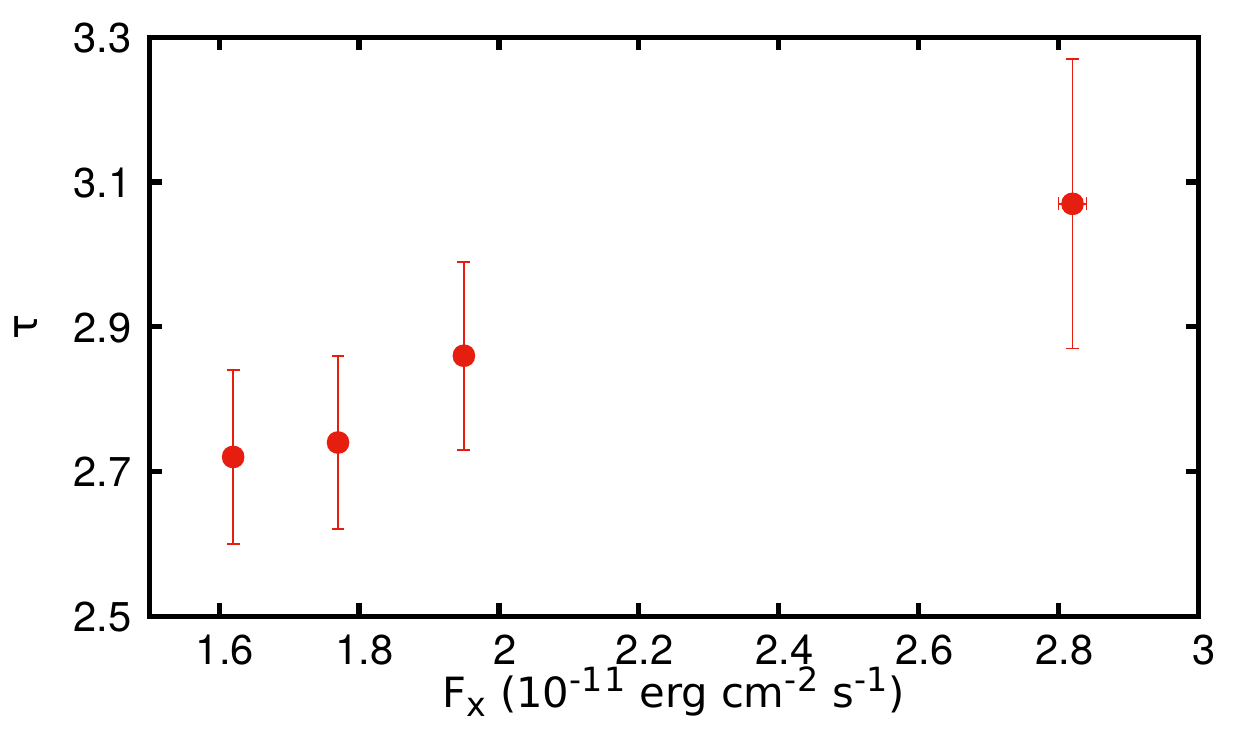}
	\caption{The variation of coronal electron temperature ($top$ panel), the photon index ($middle$ panel) and the optical depth ($bottom$ panel) with the 3--50 keV flux.} 
	\label{fig:correlation_figure}
\end{figure}

We then use the four flux levels to study the detailed flux resolved spectroscopy of
the source. We use the data in the 3--50 keV energy band and fitted FPMA and
FPMB spectra simultaneously with the same reflection model {\tt xillverCp}. The flux resolved spectra from the four levels are shown in Figure \ref{fig:4spectra}. 
We fix the iron abundance (Afe) and inclination
($\theta$) to the values obtained from the time-averaged spectrum and
kept the other parameters free. This model described
the observed spectra in different flux levels and we obtained a
satisfactory fit in all cases. We estimated the flux
in the energy range 3--50 keV using the XSPEC model {\tt cflux}. It is seen from the
flux resolved spectral fitting that the flux of Ark 564 
changed abruptly in the fourth state (highest flux state) as compared
to the other three states. The best-fit spectral parameters with errors
are tabulated in Table \ref{tab:Table2_table} and the variation of
the temperature and index with flux are shown in Figure \ref{fig:correlation_figure}. We detect a temperature drop from $\sim 17$ keV to $\sim 14$ keV when
the flux increases.

\begin{figure*}
	\includegraphics[width=0.49\textwidth]{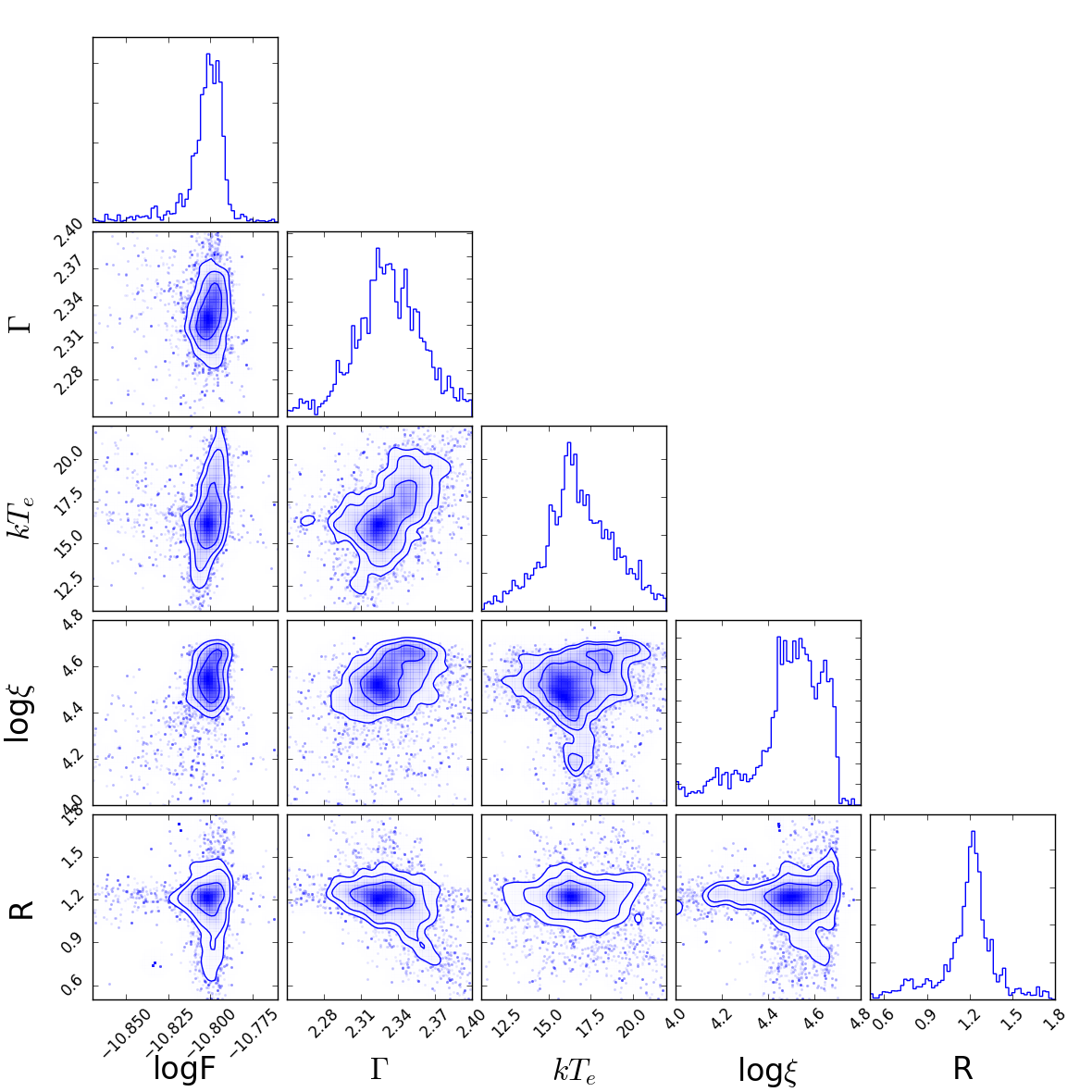}
	\includegraphics[width=0.49\textwidth]{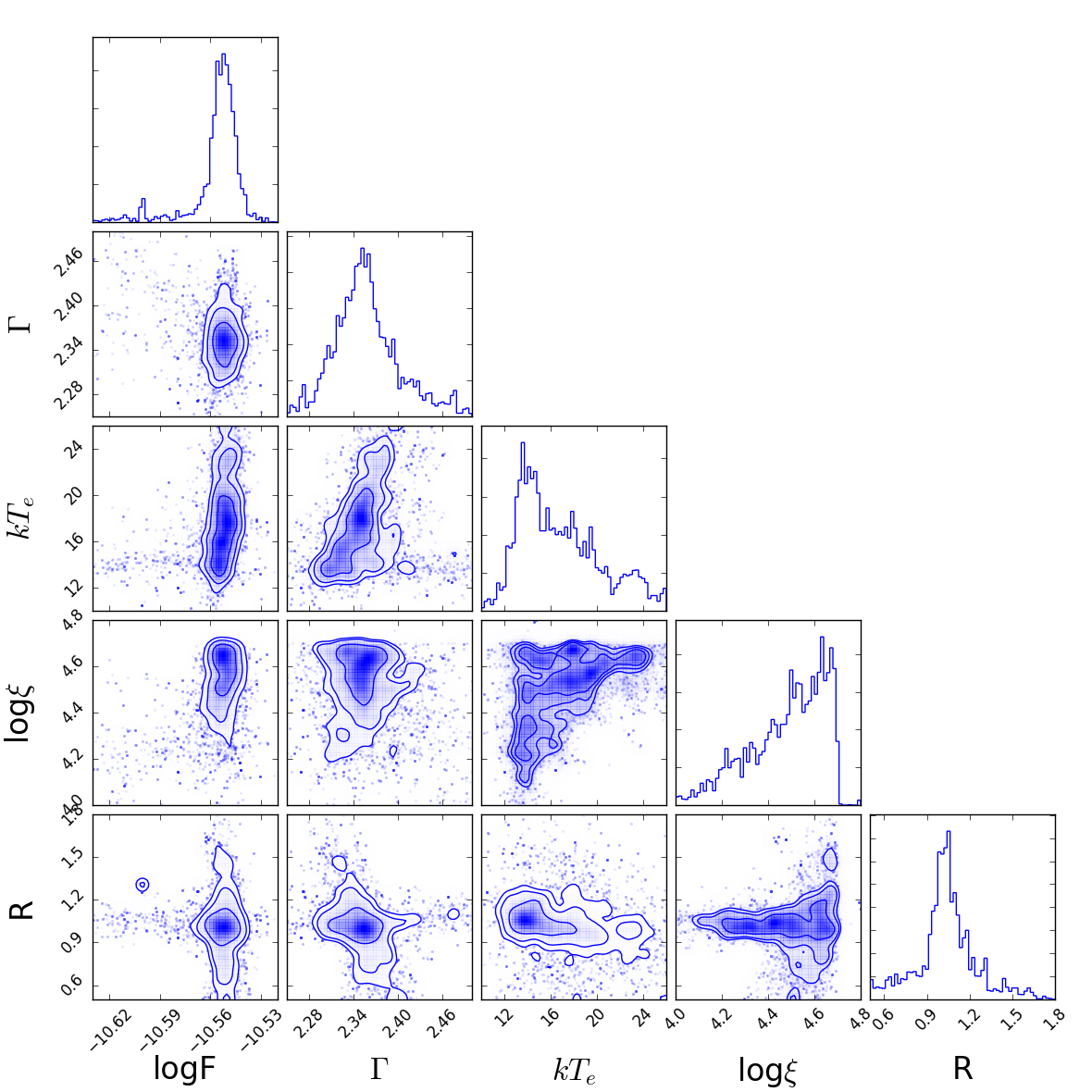}
	\caption{Corner plots of spectral parameters from MCMC analysis for the very low (left) and very high flux (right) states using \texttt{xillverCp} model. The one dimensional histograms represent the probability distribution, normalized to provide an area under the curve is equal to one. The units of log $\xi$ and 3--50 keV flux are erg cm s$^{-1}$ and erg cm$^{-2}$ s$^{-1}$, respectively.}
	\label{fig:degenplot}
\end{figure*}

To test the significance of the variation of the temperature with flux, we fitted a constant to the four data points. This resulted in a $\chi^{2}$ of 8.5 for 3 degrees of freedom, which translates to a  null hypothesis probability of $1.2\times 10^{-5}$ that the temperature did not vary with flux.
The fitting degeneracy between the spectral parameters and their
distributions are shown for the very low and very high flux states
in Figure \ref{fig:degenplot}. While the temperature is moderately degenerate with log $\xi$
and $\Gamma$, it should be noted that with flux it is only marginally correlated
and that to positively. Thus the observed anti-correlation between the temperature and flux
should not be affected by this fitting degeneracy.

The coronal temperature is statistically consistent between the 
lowest three flux bins, while  significant variation is only seen 
in the highest bin. We attempted to get a better constraint by 
binning the first three flux points into one and fitting the 
average spectrum. The best fit  temperature turned out to be 
$17.05^{+0.9}_{-0.5}$ keV, and the error bars did not significantly 
decrease compared to those obtained for the three individual flux bins.

\section{Discussion}

The flux resolved spectroscopy of Ark 564 undertaken in this work, indicates
that the coronal temperature decreases as the flux increases. For the modelling, the reflection 
fraction and the ionization parameter are allowed to vary but are not well constrained. We note that
the coronal temperature variation is detected even when these parameters are allowed to vary.

As shown in the middle panel of Figure \ref{fig:correlation_figure}, the
photon index $\Gamma$ does not vary with flux significantly. The variation in
$\Gamma$ is less than $0.1$, even though the flux varied roughly by a factor of two. For Comptonization emission, the optical depth $\tau$ is related to the
temperature $kT_e$ and photon index by \citep{Zdziarski1996,zycki1999},

\begin{eqnarray}
\tau = \sqrt{\frac{9}{4} + \frac{3}{\theta \left[\left(\Gamma+\frac{1}{2}\right)^{2} - \frac{9}{4}\right]}} - \frac{3}{2}\ \ \ \ \
\end{eqnarray}

where $\theta=kT_e/m_ec^2$. The bottom panel of Figure \ref{fig:correlation_figure} 
shows the variation of $\tau$ with flux as inferred from the above equation.
Thus, a 10\% increase in the optical depth (from $\tau \sim 2.7$ to $\tau \sim 3.1$) may 
explain why the photon index did not vary between the low and hard flux states
despite the variation in temperature. If the optical depth has remained constant
at $\sim 2.7$, the photon index would have increased to $\sim 2.6$ from $\sim 2.3$ and may have been measurable. Thus, it is possible that if the optical depth varies, then the correlation between the flux and the photon index may not that strong. In addition, the variation in the optical depth suggests intrinsic variations in the corona rather than just cooling of corona by seed UV photons. 

Like several other AGN, Ark 564 is known to have complex warm absorbers \citep{Dewangan2007, Papadakis2007, Giustini2015}, but since the energy band being considered here is 3--50 keV, these features may not affect the overall result presented here. While the {\it NuSTAR} data analysed does not require a relativistically smeared reflection component, such a complex component may affect the estimation of the coronal temperature. Apart from these caveats, the measurement of coronal temperature at different flux levels of such sources from long {\it NuSTAR} observations, will be an important step towards understanding the nature and geometry of the inner regions of
the accretion disks.

In summary, we show using flux resolved spectroscopy  of {\it NuSTAR} observation
of Ark 564,  evidence that the coronal temperature  decreases with flux.
While, the inverse correlation between the coronal  temperature and flux has been inferred
indirectly using  observed relations between spectral index and flux, this
is the first direct evidence for such a behaviour. We further infer that the
optical depth of the corona changes with flux, revealing a complex temporal
behaviour of the inner regions of the accretion flow.

\section*{Acknowledgements}
We thank the anonymous referee for the constructive comments and suggestions that improved this manuscript.
SB acknowledges IUCAA Visitors programme and thank Jermy Sanders for the detailed discussion on the {\sc xspec\_emcee} code. RS and AP acknowledge the visiting associateship programme of IUCAA. SB, RS, AP \& RM acknowledge the SERB research grant EMR/2016/005835. 
This research has made use of data obtained from the High Energy Astrophysics Science Archive Research Center (HEASARC), provided by NASA's Goddard Space Flight Center, and the {\it NuSTAR} Data Analysis Software (NUSTARDAS) jointly developed by the ASI Science Data Center (ASDC, Italy) and the California Institute of Technology (Caltech, USA).





\begin{thebibliography}{}
\makeatletter
\relax
\def\mn@urlcharsother{\let\do\@makeother \do\$\do\&\do\#\do\^\do\_\do\%\do\~}
\def\mn@doi{\begingroup\mn@urlcharsother \@ifnextchar [ {\mn@doi@}
  {\mn@doi@[]}}
\def\mn@doi@[#1]#2{\def\@tempa{#1}\ifx\@tempa\@empty \href
  {http://dx.doi.org/#2} {doi:#2}\else \href {http://dx.doi.org/#2} {#1}\fi
  \endgroup}
\def\mn@eprint#1#2{\mn@eprint@#1:#2::\@nil}
\def\mn@eprint@arXiv#1{\href {http://arxiv.org/abs/#1} {{\tt arXiv:#1}}}
\def\mn@eprint@dblp#1{\href {http://dblp.uni-trier.de/rec/bibtex/#1.xml}
  {dblp:#1}}
\def\mn@eprint@#1:#2:#3:#4\@nil{\def\@tempa {#1}\def\@tempb {#2}\def\@tempc
  {#3}\ifx \@tempc \@empty \let \@tempc \@tempb \let \@tempb \@tempa \fi \ifx
  \@tempb \@empty \def\@tempb {arXiv}\fi \@ifundefined
  {mn@eprint@\@tempb}{\@tempb:\@tempc}{\expandafter \expandafter \csname
  mn@eprint@\@tempb\endcsname \expandafter{\@tempc}}}

\bibitem[\protect\citeauthoryear{{Arnaud}}{{Arnaud}}{1996}]{Arn96}
{Arnaud} K.~A.,  1996, in {Jacoby} G.~H.,  {Barnes} J.,  eds,  Astronomical
  Society of the Pacific Conference Series Vol. 101, Astronomical Data Analysis
  Software and Systems V. p.~17

\bibitem[\protect\citeauthoryear{{Boller}, {Brandt}  \& {Fink}}{{Boller}
  et~al.}{1996}]{Boller1996}
{Boller} T.,  {Brandt} W.~N.,   {Fink} H.,  1996, \aap, \href
  {http://adsabs.harvard.edu/abs/1996A%26A...305...53B} {305, 53}

\bibitem[\protect\citeauthoryear{{Boller}, {Brandt}, {Fabian}  \&
  {Fink}}{{Boller} et~al.}{1997}]{1997MNRAS.289..393B}
{Boller} T.,  {Brandt} W.~N.,  {Fabian} A.~C.,   {Fink} H.~H.,  1997, \mn@doi
  [\mnras] {10.1093/mnras/289.2.393}, \href
  {https://ui.adsabs.harvard.edu/abs/1997MNRAS.289..393B} {289, 393}

\bibitem[\protect\citeauthoryear{{Brandt} \& {Boller}}{{Brandt} \&
  {Boller}}{1998}]{Brandt1998}
{Brandt} N.,  {Boller} T.,  1998, \mn@doi [Astronomische Nachrichten]
  {10.1002/asna.2123190104}, \href
  {http://adsabs.harvard.edu/abs/1998AN....319....7B} {319, 163}

\bibitem[\protect\citeauthoryear{{Brandt}, {Mathur}  \& {Elvis}}{{Brandt}
  et~al.}{1997}]{Brandt1997}
{Brandt} W.~N.,  {Mathur} S.,   {Elvis} M.,  1997, \mn@doi [\mnras]
  {10.1093/mnras/285.3.L25}, \href
  {http://adsabs.harvard.edu/abs/1997MNRAS.285L..25B} {285, L25}

\bibitem[\protect\citeauthoryear{{Brandt}, {Boller}, {Fabian}  \&
  {Ruszkowski}}{{Brandt} et~al.}{1999}]{1999MNRAS.303L..53B}
{Brandt} W.~N.,  {Boller} T.,  {Fabian} A.~C.,   {Ruszkowski} M.,  1999,
  \mn@doi [\mnras] {10.1046/j.1365-8711.1999.02411.x}, \href
  {https://ui.adsabs.harvard.edu/abs/1999MNRAS.303L..53B} {303, L53}

\bibitem[\protect\citeauthoryear{Buisson, Fabian  \& Lohfink}{Buisson
  et~al.}{2018}]{Buisson1093}
Buisson D. J.~K.,  Fabian A.~C.,   Lohfink A.~M.,  2018, \mn@doi [Monthly
  Notices of the Royal Astronomical Society] {10.1093/mnras/sty2609}, 481, 4419

\bibitem[\protect\citeauthoryear{{Collin} \& {Kawaguchi}}{{Collin} \&
  {Kawaguchi}}{2004}]{Collin2004}
{Collin} S.,  {Kawaguchi} T.,  2004, \mn@doi [\aap]
  {10.1051/0004-6361:20040528}, \href
  {http://adsabs.harvard.edu/abs/2004A%26A...426..797C} {426, 797}

\bibitem[\protect\citeauthoryear{{Dewangan}, {Griffiths}, {Dasgupta}  \&
  {Rao}}{{Dewangan} et~al.}{2007}]{Dewangan2007}
{Dewangan} G.~C.,  {Griffiths} R.~E.,  {Dasgupta} S.,   {Rao} A.~R.,  2007,
  \mn@doi [\apj] {10.1086/523683}, \href
  {http://adsabs.harvard.edu/abs/2007ApJ...671.1284D} {671, 1284}

\bibitem[\protect\citeauthoryear{{Fabian}, {Lohfink}, {Kara}, {Parker},
  {Vasudevan}  \& {Reynolds}}{{Fabian} et~al.}{2015}]{Fab15}
{Fabian} A.~C.,  {Lohfink} A.,  {Kara} E.,  {Parker} M.~L.,  {Vasudevan} R.,
  {Reynolds} C.~S.,  2015, \mn@doi [\mnras] {10.1093/mnras/stv1218}, \href
  {https://ui.adsabs.harvard.edu/abs/2015MNRAS.451.4375F} {451, 4375}

\bibitem[\protect\citeauthoryear{{Garc{\'{\i}}a}, {Dauser}, {Steiner},
  {McClintock}, {Keck}  \& {Wilms}}{{Garc{\'{\i}}a} et~al.}{2015}]{Gar15}
{Garc{\'{\i}}a} J.~A.,  {Dauser} T.,  {Steiner} J.~F.,  {McClintock} J.~E.,
  {Keck} M.~L.,   {Wilms} J.,  2015, \mn@doi [\apjl]
  {10.1088/2041-8205/808/2/L37}, \href
  {https://ui.adsabs.harvard.edu/abs/2015ApJ...808L..37G} {808, L37}

\bibitem[\protect\citeauthoryear{{George} \& {Fabian}}{{George} \&
  {Fabian}}{1991}]{George1991}
{George} I.~M.,  {Fabian} A.~C.,  1991, \mn@doi [\mnras]
  {10.1093/mnras/249.2.352}, \href
  {http://adsabs.harvard.edu/abs/1991MNRAS.249..352G} {249, 352}

\bibitem[\protect\citeauthoryear{{Giustini}, {Turner}, {Reeves}, {Miller},
  {Legg}, {Kraemer}  \& {George}}{{Giustini} et~al.}{2015}]{Giustini2015}
{Giustini} M.,  {Turner} T.~J.,  {Reeves} J.~N.,  {Miller} L.,  {Legg} E.,
  {Kraemer} S.~B.,   {George} I.~M.,  2015, \mn@doi [\aap]
  {10.1051/0004-6361/201425280}, \href
  {http://adsabs.harvard.edu/abs/2015A%26A...577A...8G} {577, A8}

\bibitem[\protect\citeauthoryear{{Haardt}, {Maraschi}  \&
  {Ghisellini}}{{Haardt} et~al.}{1997}]{Haardt1997}
{Haardt} F.,  {Maraschi} L.,   {Ghisellini} G.,  1997, \mn@doi [\apj]
  {10.1086/303656}, \href {http://adsabs.harvard.edu/abs/1997ApJ...476..620H}
  {476, 620}

\bibitem[\protect\citeauthoryear{{Harrison} et~al.,}{{Harrison}
  et~al.}{2013}]{Harrison2013}
{Harrison} F.~A.,  et~al., 2013, \mn@doi [\apj] {10.1088/0004-637X/770/2/103},
  \href {http://adsabs.harvard.edu/abs/2013ApJ...770..103H} {770, 103}

\bibitem[\protect\citeauthoryear{{Huchra}, {Vogeley}  \& {Geller}}{{Huchra}
  et~al.}{1999}]{Huchra1999}
{Huchra} J.~P.,  {Vogeley} M.~S.,   {Geller} M.~J.,  1999, \mn@doi [\apjs]
  {10.1086/313194}, \href {http://adsabs.harvard.edu/abs/1999ApJS..121..287H}
  {121, 287}

\bibitem[\protect\citeauthoryear{{Kara}, {Garc{\'{\i}}a}, {Lohfink}, {Fabian},
  {Reynolds}, {Tombesi}  \& {Wilkins}}{{Kara} et~al.}{2017}]{Kara2017}
{Kara} E.,  {Garc{\'{\i}}a} J.~A.,  {Lohfink} A.,  {Fabian} A.~C.,  {Reynolds}
  C.~S.,  {Tombesi} F.,   {Wilkins} D.~R.,  2017, \mn@doi [\mnras]
  {10.1093/mnras/stx792}, \href
  {http://adsabs.harvard.edu/abs/2017MNRAS.468.3489K} {468, 3489}

\bibitem[\protect\citeauthoryear{{Leighly}}{{Leighly}}{1999a}]{Leighly1999}
{Leighly} K.~M.,  1999a, \mn@doi [\apjs] {10.1086/313277}, \href
  {http://adsabs.harvard.edu/abs/1999ApJS..125..297L} {125, 297}

\bibitem[\protect\citeauthoryear{{Leighly}}{{Leighly}}{1999b}]{Leighly1999b}
{Leighly} K.~M.,  1999b, \mn@doi [\apjs] {10.1086/313287}, \href
  {http://adsabs.harvard.edu/abs/1999ApJS..125..317L} {125, 317}

\bibitem[\protect\citeauthoryear{{Papadakis}, {Brinkmann}, {Page}, {McHardy}
  \& {Uttley}}{{Papadakis} et~al.}{2007}]{Papadakis2007}
{Papadakis} I.~E.,  {Brinkmann} W.,  {Page} M.~J.,  {McHardy} I.,   {Uttley}
  P.,  2007, \mn@doi [\aap] {10.1051/0004-6361:20065527}, \href
  {http://adsabs.harvard.edu/abs/2007A%26A...461..931P} {461, 931}

\bibitem[\protect\citeauthoryear{{Pounds}, {Done}  \& {Osborne}}{{Pounds}
  et~al.}{1995}]{1995MNRAS.277L...5P}
{Pounds} K.~A.,  {Done} C.,   {Osborne} J.~P.,  1995, \mn@doi [\mnras]
  {10.1093/mnras/277.1.L5}, \href
  {https://ui.adsabs.harvard.edu/abs/1995MNRAS.277L...5P} {277, L5}

\bibitem[\protect\citeauthoryear{{Remillard}, {Grossan}, {Bradt}, {Ohashi}  \&
  {Hayashida}}{{Remillard} et~al.}{1991}]{1991Natur.350..589R}
{Remillard} R.~A.,  {Grossan} B.,  {Bradt} H.~V.,  {Ohashi} T.,   {Hayashida}
  K.,  1991, \mn@doi [\nat] {10.1038/350589a0}, \href
  {https://ui.adsabs.harvard.edu/abs/1991Natur.350..589R} {350, 589}

\bibitem[\protect\citeauthoryear{{Sarma}, {Tripathi}, {Misra}, {Dewangan},
  {Pathak}  \& {Sarma}}{{Sarma} et~al.}{2015}]{sarma2015}
{Sarma} R.,  {Tripathi} S.,  {Misra} R.,  {Dewangan} G.,  {Pathak} A.,
  {Sarma} J.~K.,  2015, \mn@doi [\mnras] {10.1093/mnras/stv005}, \href
  {http://adsabs.harvard.edu/abs/2015MNRAS.448.1541S} {448, 1541}

\bibitem[\protect\citeauthoryear{{Sobolewska} \& {Papadakis}}{{Sobolewska} \&
  {Papadakis}}{2009}]{Sobolewska2009}
{Sobolewska} M.~A.,  {Papadakis} I.~E.,  2009, \mn@doi [\mnras]
  {10.1111/j.1365-2966.2009.15382.x}, \href
  {http://adsabs.harvard.edu/abs/2009MNRAS.399.1597S} {399, 1597}

\bibitem[\protect\citeauthoryear{{Sunyaev} \& {Truemper}}{{Sunyaev} \&
  {Truemper}}{1979}]{1979Natur.279..506S}
{Sunyaev} R.~A.,  {Truemper} J.,  1979, \mn@doi [\nat] {10.1038/279506a0},
  \href {http://adsabs.harvard.edu/abs/1979Natur.279..506S} {279, 506}

\bibitem[\protect\citeauthoryear{{Turner}, {George}  \& {Nandra}}{{Turner}
  et~al.}{1998}]{Turner1998}
{Turner} T.~J.,  {George} I.~M.,   {Nandra} K.,  1998, \mn@doi [\apj]
  {10.1086/306434}, \href {http://adsabs.harvard.edu/abs/1998ApJ...508..648T}
  {508, 648}

\bibitem[\protect\citeauthoryear{{Turner}, {George}  \& {Netzer}}{{Turner}
  et~al.}{1999}]{1999ApJ...526...52T}
{Turner} T.~J.,  {George} I.~M.,   {Netzer} H.,  1999, \mn@doi [\apj]
  {10.1086/307995}, \href
  {https://ui.adsabs.harvard.edu/abs/1999ApJ...526...52T} {526, 52}

\bibitem[\protect\citeauthoryear{{Turner}, {Romano}, {George}, {Edelson},
  {Collier}, {Mathur}  \& {Peterson}}{{Turner} et~al.}{2001}]{Turner2001}
{Turner} T.~J.,  {Romano} P.,  {George} I.~M.,  {Edelson} R.,  {Collier} S.~J.,
   {Mathur} S.,   {Peterson} B.~M.,  2001, \mn@doi [\apj] {10.1086/323232},
  \href {http://adsabs.harvard.edu/abs/2001ApJ...561..131T} {561, 131}

\bibitem[\protect\citeauthoryear{{Vaughan}, {Reeves}, {Warwick}  \&
  {Edelson}}{{Vaughan} et~al.}{1999a}]{Vaughan1999b}
{Vaughan} S.,  {Reeves} J.,  {Warwick} R.,   {Edelson} R.,  1999a, \mn@doi
  [\mnras] {10.1046/j.1365-8711.1999.02811.x}, \href
  {http://adsabs.harvard.edu/abs/1999MNRAS.309..113V} {309, 113}

\bibitem[\protect\citeauthoryear{{Vaughan}, {Reeves}, {Warwick}  \&
  {Edelson}}{{Vaughan} et~al.}{1999b}]{Vaughan1999}
{Vaughan} S.,  {Reeves} J.,  {Warwick} R.,   {Edelson} R.,  1999b, \mn@doi
  [\mnras] {10.1046/j.1365-8711.1999.02811.x}, \href
  {http://adsabs.harvard.edu/abs/1999MNRAS.309..113V} {309, 113}

\bibitem[\protect\citeauthoryear{{Wilkins} \& {Gallo}}{{Wilkins} \&
  {Gallo}}{2015}]{2015MNRAS.449..129W}
{Wilkins} D.~R.,  {Gallo} L.~C.,  2015, \mn@doi [\mnras]
  {10.1093/mnras/stv162}, \href
  {https://ui.adsabs.harvard.edu/abs/2015MNRAS.449..129W} {449, 129}

\bibitem[\protect\citeauthoryear{{Zdziarski}, {Johnson}  \&
  {Magdziarz}}{{Zdziarski} et~al.}{1996}]{Zdziarski1996}
{Zdziarski} A.~A.,  {Johnson} W.~N.,   {Magdziarz} P.,  1996, \mn@doi [\mnras]
  {10.1093/mnras/283.1.193}, \href
  {http://adsabs.harvard.edu/abs/1996MNRAS.283..193Z} {283, 193}

\bibitem[\protect\citeauthoryear{{Zdziarski}, {Lubi{\'n}ski}, {Gilfanov}  \&
  {Revnivtsev}}{{Zdziarski} et~al.}{2003}]{Zdziarski2003}
{Zdziarski} A.~A.,  {Lubi{\'n}ski} P.,  {Gilfanov} M.,   {Revnivtsev} M.,
  2003, \mn@doi [\mnras] {10.1046/j.1365-8711.2003.06556.x}, \href
  {http://adsabs.harvard.edu/abs/2003MNRAS.342..355Z} {342, 355}

\bibitem[\protect\citeauthoryear{{{\.Z}ycki}, {Done}  \& {Smith}}{{{\.Z}ycki}
  et~al.}{1999}]{zycki1999}
{{\.Z}ycki} P.~T.,  {Done} C.,   {Smith} D.~A.,  1999, \mn@doi [\mnras]
  {10.1046/j.1365-8711.1999.02885.x}, \href
  {http://adsabs.harvard.edu/abs/1999MNRAS.309..561Z} {309, 561}

\makeatother
\end{thebibliography}








\bsp	
\label{lastpage}
\end{document}